\documentclass{svproc}

\usepackage{caption}
\usepackage {graphicx}
\usepackage{multicol}
\usepackage{footmisc}
\usepackage {array}
\usepackage {booktabs}
\usepackage {pseudocode}
\usepackage {algorithm}
\usepackage {algpseudocode}
\usepackage {pgfplots}
\usepackage {filecontents}
\usepackage {xcolor,listings}
\usepackage {textcomp}
\usepackage {csvsimple}
\usepackage{epstopdf}
\usepackage{hyperref}
\usepackage{orcidlink}

\newcommand\copyrighttext{%
  \newline
  \footnotesize 
  \centering
  
  Khan, A., Ghosh, S., Ghosh, S.K. (2018). eDWaaS: A Scalable Educational Data Warehouse as a Service. In: Abraham, A., Muhuri, P., Muda, A., Gandhi, N. (eds) Intelligent Systems Design and Applications. ISDA 2017. Advances in Intelligent Systems and Computing, vol 736. Springer, Cham.
  
  \url{https://doi.org/10.1007/978-3-319-76348-4\_96}
  
  \copyright 2018 Springer International Publishing AG, part of Springer Nature
  }
\newcommand\copyrightnotice{%
\begin{tikzpicture}[remember picture,overlay]
\node[anchor=south,yshift=25pt] at (current page.south) 
{{\parbox{\dimexpr\textwidth-\fboxsep-\fboxrule\relax}{\rule{\textwidth}{1pt}\copyrighttext}}};
\end{tikzpicture}%
}

\begin{document}
\mainmatter
\title{eDWaaS: A Scalable Educational Data Warehouse as a Service}
\titlerunning{eDWaaS: A Scalable Educational Data Warehouse as a Service}

\author{Anupam Khan \and Sourav Ghosh \and Soumya K. Ghosh}
\authorrunning{Anupam Khan \and Sourav Ghosh \and Soumya K. Ghosh}
\institute{Department of Computer Science and Engineering \\Indian Institute of Technology Kharagpur, India 721302 \\
	\email{anupamkh@iitkgp.ac.in, sourav.ghosh@iitkgp.ac.in \orcidlink{0000-0003-1866-1408}, skg@iitkgp.ac.in}
}

\maketitle

\copyrightnotice

\begin{abstract}
	The university management is perpetually in the process of innovating policies to improve the quality of service. Intellectual growth of the students, the popularity of university are some of the major areas that management strives to improve upon. Relevant historical data is needed in support of taking any decision. Furthermore, providing data to various university ranking frameworks is a frequent activity in recent years. The format of such requirement changes frequently which requires efficient manual effort. Maintaining a data warehouse can be a solution to this problem. However, both in-house and outsourced implementation of a dedicated data warehouse may not be a cost-effective and smart solution. This work proposes an educational data warehouse as a service (eDWaaS) model to store historical data for multiple universities. The proposed multi-tenant schema facilitates the universities to maintain their data warehouse in a cost-effective solution. It also addresses the scalability issues in implementing such data warehouse as a service model.
\end{abstract}


\graphicspath{ {supportings/} }

\section{Introduction}
	Higher education is a public welfare provided by nonprofit organisations with a  societal mission. However, nowadays it is becoming a global service in an ever-more complex and competitive knowledge marketplace. To survive, the higher educational universities started prioritising revenue earning and promoting the university at global level. In addition to that, sustaining reputation has become the primary focus of the management. Competition in academia is the driving force that influences the universities to always innovate policy for enhancing the quality of service \cite{pucciarelli2016competition}. In this age of information, several ranking framework help a university to maintain a better visibility globally. In addition to global reputation, management also tries to increase popularity of the departments, specialisation, courses etc. 
	\par
	The decision makers of a university often need historical data and its' analysis for framing future policies and providing information to several ranking frameworks. Unfortunately, these required data, especially the historical information, are not easily available. A data warehouse can be an effective solution to store the required historical data after extract-transform-load (ETL) operation of information from heterogeneous sources. The online analytical processing (OLAP) of queries help in generating report easily in desired format. In fact, there is a pressing demand of designing a data warehouse for the universities \cite{idris2014framework}. Moreover, it can facilitate the educational data mining, which is becoming popular nowadays \cite{khan2016analysing}. However, it is a costly affair to maintain such data warehouse in-house. Similarly, outsourcing the dedicated data warehouse maintenance to third-party organisation is not a smart solution. A dedicated data warehouse usually requires high-end hardware for complex information retrieval but may not yield healthy utilisation out of it.
	\par
	The data warehouse as a service (DWaaS) model has emerged as a growing market in recent years, but there is little significant advancement in this field so far. To best of our knowledge, the term DWaaS was first coined in 2012 \cite{kaur2012visualizing}. Though, the performance issues due to large data volume, heavy duty ETL process are still matter of concern \cite{agrawal2017best}. Some researchers have even tried to implement ETL and OLAP processes on cloud platform \cite{saada2011cloud,cuzzocrea2014cloud}. However, the scalability issues are not clearly addressed by them. Some researchers have even proposed various data marts suitable in education domain \cite{dell2007academic,idris2014framework,di2015academic,kurniawan2013use}. However, a robust implementation of DWaaS in education domain is not clearly identifiable in literature. Therefore, the aim of this work is as follows: (i) designing a multi-tenant schema for universities in cloud environment, and (ii) implementing a scalable ETL and OLAP process.
	\par
	This study proposes an educational data warehouse as a service (eDWaaS) model offered in multi-tenant environment. It helps the universities to maintain their historical statistics in a cost effective environment. The subscriber gains by offering the service to multiple universities; thus increasing the utilisation of resources. The multi-tenant architecture reduces the operating cost for the service consumer. The proposed approach to ETL process and report generation using OLAP query handles the scalability issues in cloud data warehouse implementation.
	\par
	Section \ref{sec:edwaas} describes the proposed eDWaaS model. Section \ref{sec:isitscalable} presents the analysis on the scalability of the proposed model. Finally, section \ref{conclusion} summarises this work and highlights the future direction.
	\begin{figure}
		\centering
		\includegraphics[width=12.0cm]{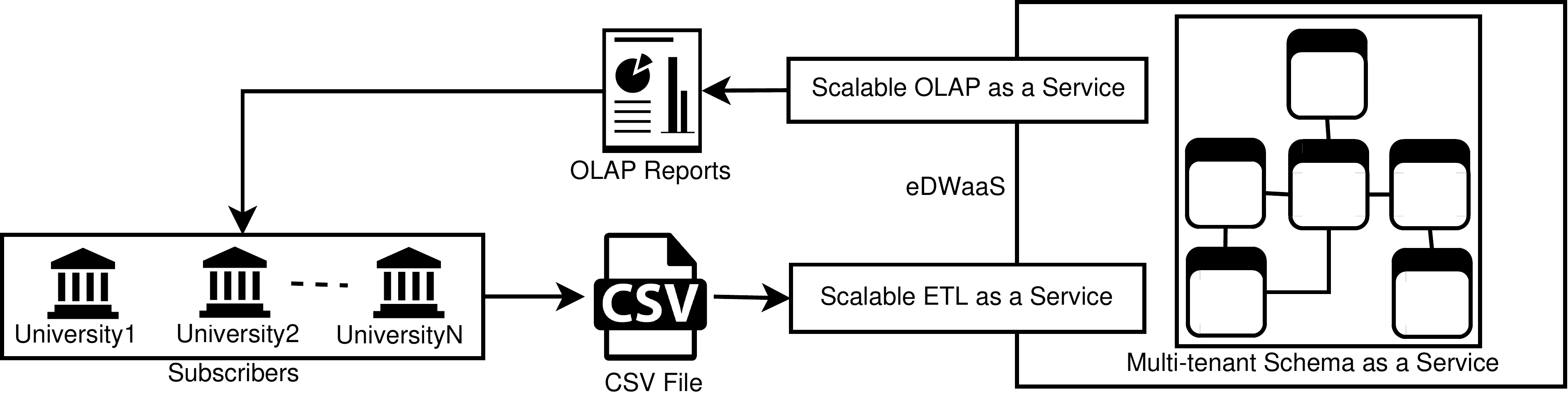}
		\caption{Architecture of eDWaaS}
		\label{fig:arch}
	\end{figure}
	
	\section{eDWaaS Model}  \label{sec:edwaas}
	This section elaborates the proposed eDWaaS model for building up the academic data repository. Figure \ref{fig:arch} presents the architecture of multi-tenant eDWaaS model that enable multiple universities to subscribe the services provided by it. The multi-tenant schema on Hive is the central component of eDWaaS. The ETL service felicitates the university to upload the data in widely accepted comma separated value (CSV) format. The OLAP service produces report from the multi-tenant schema.

	\subsection{Schema as a Service}  \label{subsec:scaas}
	As a core of the eDWaaS model, it provides a multi-tenant schema of academic data mart. It is important to mention that the proposed schema does not consider all aspects of academic data mart. Rather, it is only a representative one which considers the following few aspects among them: (i) student performance (ii) teaching quality, and (iii) student counts. 
	
	\begin{figure*}[!hbt]
		\centering
		\includegraphics[width=12.0cm]{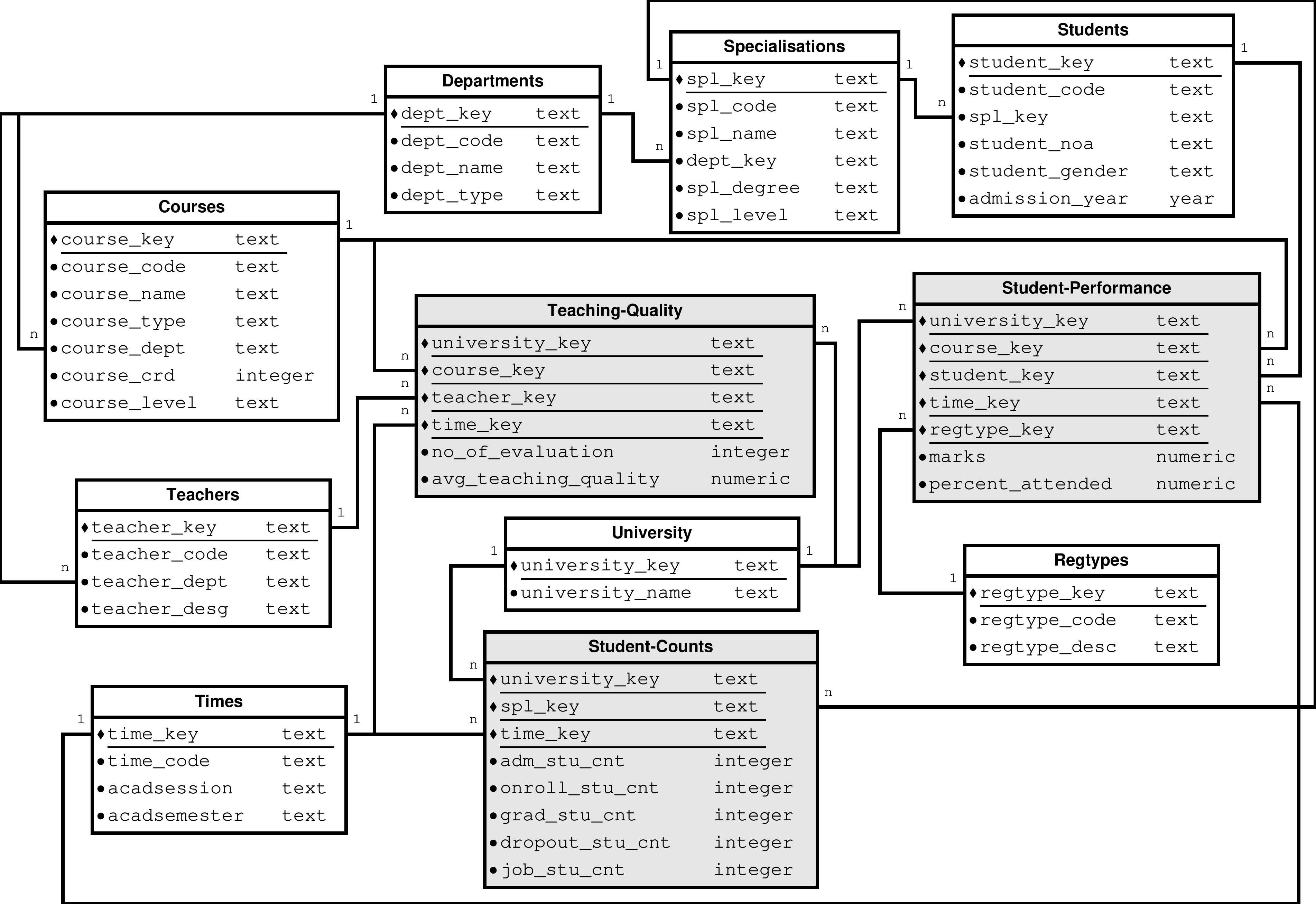}
		\caption{Multi-tenant schema for academic data mart}
		\label{fig:schema-cloud}
	\end{figure*}
	
	The snowflake schema of this data mart is designed in such a way that it can efficiently manage information for multiple universities. The proposed schema, containing dimensions and fact tables, represents a multi-tenant model which ensures the data abstraction between multiple universities. The dimension of academic data mart enables the consumer to categorise various interesting measures. The proposed schema contains eight dimensions and three fact tables. Figure \ref{fig:schema-cloud} presents these fact tables in grey background and dimension tables in white background. 
	\par
	In the proposed snowflake schema, unification of records in dimension table is handled by prefixing university key with the dimension key. For example, the dimension key of \textit{student1} of \textit{university1} in \textit{Students} table is \textit{university1\_student1}. The fact table maintains quantifiable measure of certain attributes based on different dimensions. In addition to that, the proposed schema maintains the university key as a separate attribute in each fact table to ease out the process of filtering data during OLAP queries. For example, the \textit{university\_key} in \textit{Student-Performance} table helps in this filtration process. 
	
	\subsection{ETL as a Service} \label{subsec:etl-service}
	
\begin{figure*}
		\centering
		\includegraphics[width=12.3cm]{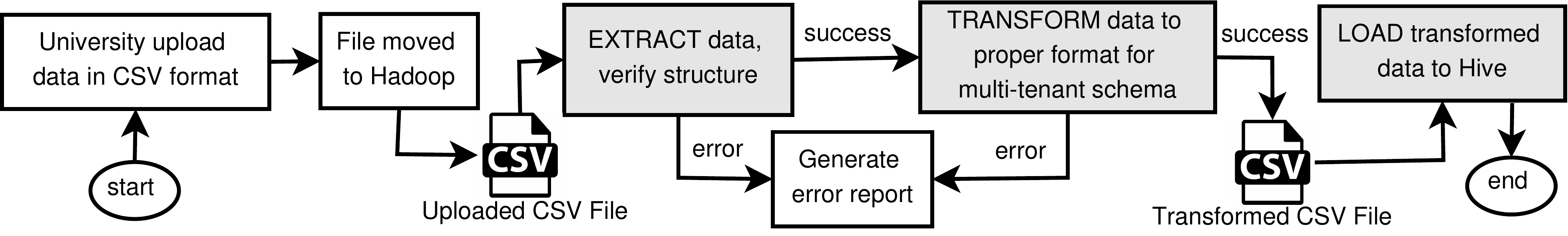}
		\caption{Overview of ETL process}
		\label{fig:ETL-overview}
	\end{figure*}

	In this multi-tenant model, growth of data and increasing amount of simultaneous requests are the two challenging issues for the DWSP. From the perspective of a university, performance of the data warehouse should be independent of the activities of other tenants. Therefore, the next concern is to design the data warehouse related processes in a manner which would scale well with increasing data and request. Transferring the data to the multi-tenant cloud data warehouse through ETL extract-transform process is a difficult task. The simplicity of this process can make the eDWaaS model successful. In this approach, the universities upload the student record using the ETL process offered as a service. The ETL service converts the uploaded data to make it suitable for storing in the multi-tenant schema. An overview of the ETL process is presented in Figure \ref{fig:ETL-overview} and discussed in the following part of this section.
	\par
	The universities upload the dimension and fact data in CSV format from their tenant login. Complex operations are handled by the data transformation process. One of such complex operation is the dimension key generation. Let us consider the example of \textit{student\_key} in \textit{Dim-Students} table. In this implementation, the subscribed university uploads any of their specific student identifier like roll number, enrolment id, registration number etc. The ETL process generates the \textit{student\_key} by concatenating \textit{university\_key} with uploaded student identifier. This helps the processing to be simple. It also helps the DWSP to maintain identity of the records in multi-tenant schema. The process involved in transferring data from university sources to the multi-tenant schema is collectively referred to as ETL as a service. In order to make these process scalable, this study has exploited Map-Reduce framework on a Hadoop cluster for extract and transform process. 
	
	\subsubsection{Extract Process}
	
	This study uses a predefined structure of CSV data which the university has to follow. The system detects any deviation from such format by comparing the CSV structure with the acceptable input format. The process aborts once in case of any deviation. In such cases, the system generates a report mentioning the line in the CSV file where the error has been observed. Otherwise, the process continues, parsing lines from the CSV file one at a time, and storing the data in an intermediate file. In this implementation, several mappers consume the lines from the CSV file and verify the structure of the data. The proposed model left the data cleansing, the process of detecting and rectifying erroneous and/or inconsistent records, as a manual process to the subscriber. 
	
	\subsubsection{Transform Process}
	The transformation process converts the extracted data in the proposed schema acceptable format. Since the data from multiple universities are stored in a multi-tenant schema, the transformation process generates dimension keys by concatenating the \textit{university\_key} with the key provided by the university in the CSV input. An important concern here is whether to store only the transformed key as a single attribute; or to store the tenant provided key and the \textit{university\_key} as separate attributes in dimension tables. The former approach allows faster join operations in reporting. However, it requires some post-processing overhead in report generation. Similarly, the second approach implies a concatenation overhead in OLAP queries. Thus, the former approach achieves performance benefits during query execution at the cost of overhead in report generation. The second approach does the opposite.
	\par
	This work adapts a third approach in this implementation, where the uploaded key of a dimension is stored along with the transformed key value. This sacrifices linear fraction of storage space but decreases computation time for both OLAP query execution and report generation. In this approach, the transformation program scans each record and appends the \textit{university\_key} with all attributes which are either a key or a reference to some dimension table. In this implementation, mappers transform the data immediately after extracting from input file and store the transformed output in an intermediate file of Hadoop file system.
	
	\subsubsection{Load Process}
	The final step in ETL involves loading the transformed data in the multi-tenant schema. The primary requirement of ETL process is to ensure that the data warehouse is in a consistent state at any point of time. A secondary requirement is to conveniently generate a report of erroneous data or inconsistencies observed at this stage. The proposed ETL process discards the entire input file in case of error in data. The generated report enlists the line numbers, and tenant provided unique key values, and the nature of the error. However, in case of no such error, the load process moves the transformed data to the Hive tables in Hadoop file system (HDFS).
	\par
	Directly loading the transformed file using Hive queries would lead to faster load operation at the cost of occasional inconsistent or duplicate data. However, these can be discarded easily during report generation without much overhead. Thus, this implementation uses a simple \texttt{LOAD DATA} query to load the intermediate file of transform phase to the corresponding schema in HDFS. It is important to mention that if the HDFS and Hive are residing on the same Hadoop cluster, then this operation would involve merely moving the file from one location in HDFS to another. Hence, it requires only constant time regardless of the input file size.
	
	\subsection{OLAP as a Service} \label{subsec:olap-service}
	
	In the proposed eDWaaS model, the subscriber initiates the report generation through an online interface provided to them. The subscribers gain access to this interface by authenticating themselves with login identifier and password provided by the DWSP. It is important to mention here that the present eDWaaS model deals with some pre-defined reports only. This study facilitates the subscriber to generate those reports by processing the pre-defined data cubes at backend. The system, in turn, executes the OLAP query on the cube to extract the data required for a report. This mechanism of report generation is termed as OLAP as a service. In the proposed eDWaaS model, the logical separation of data is achieved by defining scope of the OLAP queries. The \textit{university\_key} attribute in the fact table ensures the scope to be limited to the data concerned. This study has filtered the data by subscribers' \textit{university\_key} available in the authenticated session of the web-based application software. The system then generates the report based on the filtered data only.
	
	\subsubsection{Data Cube Generation} \label{subsubsec:data-cube}
	The process of data cube generation takes a considerable time in any data warehouse implementation. In order to reduce this time, this work takes an approach where the data are processed to generate the pre-defined cubes offline. These cubes are generated from the fact tables and related dimensions. The system updates the content of the cubes at regular intervals. As a result, after some data is uploaded through ETL service, the system requires a minimum time 
	to make it available in the report
	.
	\par
	This study creates the data cube using queries of the form \texttt{CREATE TABLE ... WITH CUBE}. The Hive, in turn, creates a table where redundant data are stored from the fact and related dimensions tables. The data cube takes up more space than the collective space taken up by the related fact and the dimension tables. However, data cube enables the parallel processing of OLAP queries by the Hadoop nodes in less time with respect to the non-cube queries on the dimensions and fact tables. Joining of relevant tables is the major bottleneck of the parallel processing in the case of non-cube queries. In such cases, the non-cube query first fetches the filtered data from these individual tables and then joins them as per requirement. This join operation needs to wait until data from all individual tables have been fetched. However, in OLAP cube, parallel system can filter records by reading them independently. Thus, the query executes faster as there is no waiting time for joining.  One of such data cube is presented below.
	
	\begin{lstlisting}[
	%language=SQL,
	showspaces=false,
	basicstyle=\ttfamily \scriptsize,
	%numbers=left,
	%numberstyle=\footnotesize,
	%commentstyle=\color{gray}
	]
CREATE TABLE cube_student_performance 
AS SELECT grouping_id, F.university_key, C.course_code, T.time_code, 
  R.regtype_code, AVG(marks) AS avg_marks, AVG(percent_attended) AS avg_per_att
FROM Student-Performance F JOIN Courses C ON C.course_key = F.course_key JOIN 
  Times T ON T.time_key = F.time_key JOIN Regtypes R 
  ON R.regtype_key = F.regtype_key
GROUP BY F.university_key, C.course_code, T.time_code, R.regtype_code 
WITH CUBE;
	\end{lstlisting}
	
	The system adds the \texttt{grouping\_id} during the offline data cube generation. The attributes of data cube contains \texttt{NULL} value in many records. If an attribute has a \texttt{NULL} value, the corresponding bit in \texttt{grouping\_id} is 0; otherwise it is 1. It is important to mention here that the most significant bit is from the reverse order of attributes in the \texttt{CREATE} statement. For example, a record in the above cube with \texttt{regtype\_code} and \texttt{course\_code} only as \texttt{NULL} value, the corresponding \texttt{grouping\_id} is $ 0101 $. Similarly,  a record with only \texttt{time\_code} as \texttt{NULL}, the corresponding \texttt{grouping\_id} is $ 1011 $.
	
	\subsubsection{OLAP Query}
	The system executes an OLAP query on the cube to get the final data for reporting. One of such query based on the \texttt{cube\_student\_performance} is presented below. 
	
	\begin{lstlisting}[
	%language=SQL,
	showspaces=false,
	basicstyle=\ttfamily \scriptsize,
	%numbers=left,
	%numberstyle=\tiny,
	%commentstyle=\color{gray}
	]
SELECT time_code, regtype_code, avg_marks
FROM cube_student_performance
WHERE(grouping_id = conv("010", 2, 10) OR grouping_id = conv("110", 2, 10))
  AND university_key = 'University1' AND time_code = '2016-17-SPR';
	\end{lstlisting}
	
	In this query, the \texttt{university\_key = 'University1'} clause filters all data of \texttt{University1} from the data cube. This clause provides the logical abstraction of data in multi-tenant environment. Furthermore, the \texttt{time\_code = '2016-17-SPR'} statement filters all data of \texttt{2016-17-SPR}. The \texttt{conv} is a Hive function that converts a number in a specified base to another. For example, \texttt{conv("010", 2, 10)} converts 010 from base 2 to base 10. Therefore, the above OLAP query extracts the pre-computed aggregated values for all possibile \texttt{regtype\_code} form the OLAP cube. 
	
	\section{Scalability of eDWaaS} \label{sec:isitscalable}
	This study proposes a multi-tenant schema as a service model. It also presents the service model of two major operations in a data warehouse: ETL and OLAP. However, a cloud implementation of a data warehouse can not succeed if these operations are not scalable. The size of data warehouse grows with increasing subscription from multiple universities. The service model may fail if the ETL or OLAP process takes considerable amount of time in such situation. Therefore, this work analyses the scalability aspects of the eDWaaS. The scalability analysis is laid out in two broad directions: (i) assessing the scalability of the ETL process and evaluating the benefits over a conventional system, and (ii) analysing the performance for OLAP queries over the conventional data warehouse.
	
	\subsection{Scalability of ETL Process}
	In order to verify the scalability of the designed process, this work have configured the ETL process in two ways. \textit{Case 1}: It allocates only two mappers regardless of the input file size. \textit{Case 2}: Secondly, it allocates number of mappers which is proportional to the size of the input file. Say, $S_{ip}$ is the input file size and  $S_{b}$ is the size of each block required to store files in the HDFS. The $S_{min}$ and $S_{max}$ are configurable parameters for minimum and maximum split size on the input file. The size of the split files ($S_{split}$) can be calculated using Equation \ref{eq:ipsplitsize}.
	\begin{equation}
		S_{split} = max(S_{min},\ min(S_{max},\ S_{b}))
		\label{eq:ipsplitsize}
	\end{equation} 
	Now, by configuring $S_{min}$ and $S_{max}$ to use a constant value, this work has made the $S_{split}$ independent of $S_{b}$. This is achieved by taking the following approach. The $S_{min}$ and $S_{max}$ are two constant parameters. For this experiments, $S_{min}\ =\ S_{max}\ =\ c$, where $c$ is a constant. Therefore, the Equation \ref{eq:ipsplitsize} can be re-written as Equation \ref{eq:ipsplitsizemod}.
	
	\begin{equation}
	S_{split} = max(c,\ min(c,\ S_{b}))
	\label{eq:ipsplitsizemod}
	\end{equation} 
	
	When $(c \le S_{b})$, then $S_{split}\ =\ max(c,\ c)\ =\ c$. Again, if $(c > S_{b})$, then $S_{split}\ =\ max(c,\ S_{b})\ =\ c$. Therefore, $S_{split}$ is constant irrespective of the value of $S_{b}$. The number of mappers ($n_{m}$) can be calculated using Equation \ref{eq:noofmapper}.
	\begin{equation}
	n_{m} = \frac{S_{ip}}{S_{split}}
	\label{eq:noofmapper}
	\end{equation}
	
	Therefore, the number of mappers is directly proportional to the size of input file, if $ S_{split} $ is a constant.  For \textit{Case 1} of this analysis, this study have set $ S_{split} $ equal to the half of the $S_{ip}$. This ensures the number of mappers to be 2.	This work have prepared eleven datasets comprising data of various size like 2 MB, 4 MB, 8 MB, 16 MB, ..., 1 GB, 2 GB. Thereafter, it has performed the proposed ETL process multiple times on these datasets and recorded the time taken to perform this operation. For each of the eleven datasets, the ETL process is performed for 200 times through an automated program. Out of these 200 observations in each case, this study have removed the outliers first by dropping data points beyond the inter-quantile range of 25\%\textsuperscript{ile} - 75\%\textsuperscript{ile} in each identical input file size. Furthermore, this study has also removed data points that lie outside the inclusive range of $\mu \pm (1.5)\sigma$. Here $\mu$ is the mean time taken and $\sigma$ is the standard deviation for a certain input file. Once the outliers have been removed, this work calculates the mean time for each input file size. The same process is repeated for both \textit{Case 1} and \textit{Case 2}. Figure \ref{fig:plot-extract-transform} plots two line graph with the observed data for \textit{Case 1} and \textit{Case 2}. The observed result ensures that the time taken by ETL process increases almost linearly with input file size in \textit{Case 1}. However, in \textit{Case 2},  the time is relatively constant. In \textit{Case 2}, this study has set the number of mappers to be directly proportional to the input file size. The number of mapper increases proportionately with growing file size. It is achieved by fixing $S_{split}$ to a constant value. The processing time indicates that the proposed ETL process is scalable, subject to the fact that there are sufficient number of mapper available in the system.
	\par
	\begin{minipage}[b]{.5\linewidth}
		\begin{tikzpicture}[scale=0.6]
		\begin{axis}[
		xlabel={Input file size (in bytes)},
		ylabel={Time taken for ETL process (in ns)},
		legend pos=north west,
		ylabel near ticks
		]
		\addplot [smooth,mark=*,red] table [x=x, y=y, col sep=comma] {data/etl_data.csv};
		\addlegendentryexpanded{Case 1}
		\addplot [smooth,mark=o,blue] table [x=x, y=z, col sep=comma] {data/etl_data.csv};
		\addlegendentryexpanded{Case 2}
		\end{axis}
		\end{tikzpicture}
		\captionsetup{font=scriptsize}
		
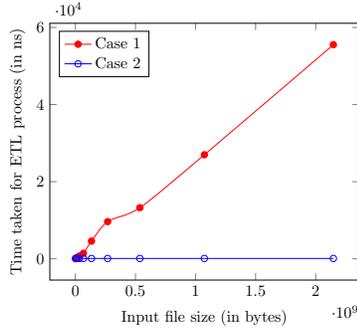
\captionof{figure}{Comparison of ETL process}
		\label{fig:plot-extract-transform}	
	\end{minipage}
	\begin{minipage}[b]{.5\linewidth}
		\begin{tikzpicture}[scale=0.6]
		\begin{axis}[
		xlabel={Records in OLAP cube},
		ylabel={Query execution time\ (in seconds)},
		legend pos=north west,
		ylabel near ticks
		]
		\addplot [smooth,mark=*,red] table [x=x, y=y, col sep=comma] {data/olap_data.csv};
		\addlegendentryexpanded{Cumulative}
		\addplot [smooth,mark=o,blue] table [x=x, y=z, col sep=comma] {data/olap_data.csv};
		\addlegendentryexpanded{Effective}
		\end{axis}
		\end{tikzpicture}
		\captionsetup{font=scriptsize}
		
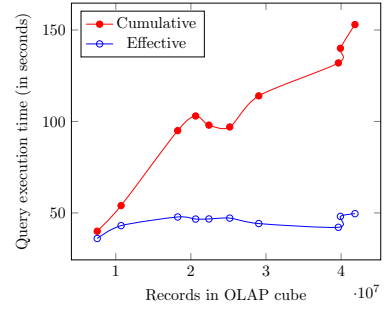
\captionof{figure}{Comparison of OLAP query}
		\label{fig:plot-olap}	
	\end{minipage}

	\subsection{Scalability of OLAP query} \label{section:olap_performance}
	In order to check the scalability of the OLAP query process, this study has analysed the proposed OLAP as a service model on the data cube mentioned in subsection \ref{subsubsec:data-cube}. The analysis considers ten data cube with varied data size starting from 7.5 to 42 million approximately. This work has executed the OLAP query over the each such cube 200 times and measured the mean effective time taken to fetch the results. It also records the mean cumulative time taken by all the nodes. The mean cumulative time indicates the effective time where the number of processing nodes is constant. The observed mean effective and cumulative time taken for the OLAP query are presented in Figure \ref{fig:plot-olap}.
	\par
	The plot in Figure \ref{fig:plot-olap} indicates that the cumulative query execution time increases with growing data size in OLAP cube. However, the effective query execution time remains almost constant. Therefore, it can maintain the effective OLAP query execution time constant by increasing number of mappers. 
	
	\section{Conclusion} \label{conclusion}
	The proposed eDWaaS model plays an important role in retrieving aggregate data efficiently from large dataset. In the present context, the proposed eDWaaS model helps the university management to generate report on student and teacher performance, various statistics on placement activity, performance of department and specialisation etc. This study presents a scalable data warehouse as a service model which can cater to the needs of multiple universities. The proposed model involves a data warehouse service provider which improves the economy of scale. This study has presented a multi-tenant schema to facilitate the service of data aggregation and analysis. In addition to this, it also proposes the scalable service model for uploading data in multi-tenant schema and extracting reports from it. The scalability of the ETL process and the performance of OLAP query have been verified with theoretical justification. 
	\par
	The proposed model uses a pre-defined schema which may not be suitable for all universities. They may need a customized schema that serves other individual requirements. Therefore, further research can be carried out to design a generalized schema which can accommodate variation in data structure. Data security in this multi-tenant schema is also an interesting area which can be explored further.

\bibliographystyle{spmpsci}
\bibliography{library}

\end{document}